\documentclass[aps,prc,groupedaddress,showkeys]{revtex4-2}
\usepackage{color}
\usepackage{graphicx}

\usepackage[dvipsnames]{xcolor}

\usepackage{cancel}

\newcommand{\beq}{\begin{equation}}
\newcommand{\eeq}{\end{equation}}

\usepackage[reqno]{amsmath}
\usepackage{amsthm}
\usepackage{amssymb}

\usepackage{gensymb}

\begin{document}

\title{Deblurring for Nuclei: 3D Characteristics of Heavy-Ion Collisions}

\author{Pawel Danielewicz}
\email[]{daniel@frib.msu.edu}
\affiliation{Facility for Rare Isotope Beams and
Department of Physics and Astronomy, Michigan State University, East Lansing, Michigan 48824, USA}

\author{Mizuki Kurata-Nishimura}
\affiliation{RIKEN Nishina Center, Hirosawa 2-1, Wako, Saitama 351-0198, Japan}

\date{\today}

\begin{abstract}

Observables from nuclear and high-energy experiments can be degraded by detector performance and/or methodology in extracting the observables, such as of the final-state characteristics of heavy-ion collisions in relation to a coarsely estimated reaction-plane direction.  We propose the use of deblurring methods, such as in optics, to correct for observable degradation.  Our main focus is the restoration of triple-differential particle distributions in heavy-ion collisions.  We demonstrate that these could be extracted from collision measurements following the Richardson-Lucy deblurring method from optics.  We illustrate basic features of the restoration methodology in a schematic model assuming either ideal or more realistic particle detection.  The inferred three-dimensional (3D) distributions for collisions may easier to interpret in terms of collision dynamics and sought properties of bulk matter than the currently employed Fourier coefficients, that combine information from different azimuthal angles relative to the reaction plane.

\end{abstract}


\maketitle

\section{Introduction}
Observables from nuclear and high-energy experiments can be degraded by detector performance and/or methodology in extracting the observables, such as tied to limited control over the initial state of a reaction. With the latter respect, the three-dimensional (3D) particle distributions relative to the reaction plane in heavy-ion collisions are of much interest, but the reaction plane direction can be only coarsely assessed, so effectively only blurred distributions could be directly determined.  The characteristics at a fixed impact parameter are of interest for the heavy-ion collisions, but again the impact parameter can be assessed only coarsely.  Direct reactions of fast rare isotopes are of much interest, but fast rare isotopes are produced in flight with their beam consequently spread out in space, angle and energy, potentially erasing details in observables.  There may be different strategies to deal with such situations, such as to regress in the observables, particularly to those that are less degraded by experimental limitations or easier to correct.  Else, one can resort to theory and filter the theoretical predictions through the features of the experimental setup and methodology.  The problem there is that nuclear theory is usually approximate at one level or another.  With observables being degraded and being compared against uncertain reference, important insights might get lost.  As an alternative to the practice so far, we propose the use of deblurring techniques to cope with any degraded observables.  The deblurring has a long record in optics, being employed there to correct images for flaws of optical apparatus. With the deblurring as an option, one may have a wider range of observables to choose from and may let those observables to stand on their own, prior to any theory comparisons, inspiring the latter rather than being resigned to them.  Of our primary interest is the inference of 3D characteristics of the final states of heavy-ion collisions.  However, different strategy details may be relevant for other contexts in nuclear physics and potentially in high-energy physics.

In heavy-ion collisions, any directional aspects of observables have been primarily tied to the incident beam direction, as in particular evidenced in the use of rapidity along the beam and/or transverse momentum magnitude in presenting data.  However, the symmetry around the beam direction is broken in the initial state by the relative displacement of the nuclear centers, by impact parameter. The transverse displacement and the beam axis define the reaction plane for the collision. The symmetry breaking has important consequences for the collisions, such as in setting the preferred direction of transverse momentum transfer in the system and in the emergence of participant and spectator products in more energetic collisions.  At lower energies, heavy projectile and target residues emerge from a collision largely aligned with the reaction plane and other characteristics of the final state tend to be related to these.  At higher energies, much could be learned on reaction of strongly interacting matter to compression, from product emission in correlation to the reaction plane.  However, at the same time, estimating the reaction plane becomes a challenge at those higher energies since the system largely vaporizes.  Our considerations here will largely refer to the latter situation.  At still higher energies, Lorentz dilation freezes fluctuations present the initial state, that can then compete with the breaking of the symmetry around the beam axis by the reaction plane.  Even there our proposed methodology may have applications, though it might not provide the most interesting information one may seek in the context.  

Importantly, the way in which the symmetry breaking in the initial state gets reflected in transverse anisotropies in particle emission, that may represent transverse collective motion and shadowing, can provide significant information about the colliding system, including temporal progress of the collision and bulk properties of the matter such as equation of state and viscosity.  For that reason, strategies not only to identify the reaction plane direction, but also quantify the azimuthal anisotropies have been intensely pursued over decades, starting with telling whether the transverse collective motion is there at all \cite{gyulassy_nuclei_1982,danielewicz_jacobian_1983}, through estimation of the reaction plane direction and its uncertainty~\cite{danielewicz_transverse_1985,ollitrault_anisotropy_1992}, to detailed quantification of anisotropies in terms of Cartesian moments~\cite{danielewicz_transverse_1985, danielewicz_collective_1988} and eventually Fourier coefficients~\cite{demoulins_measurement_1990, *demoulins_collective_1989, voloshin_flow_1996, poskanzer_methods_1998,borghini_flow_2001,bilandzic_flow_2011}.  Importance of correcting for detector inefficiencies and circumventing correlations unrelated to spatial anisotropies, that could distort inferences, was recognized  \cite{danielewicz_collective_1988,borghini_flow_2001,bilandzic_flow_2011}.

We will be connecting to those prior developments here at different level, whether directly, when addressing heavy-ion collisions, or in inspiration, behind the considerations in a simplified model.  Of specific importance is that  correlation of particle emission pattern with the reaction plane can be exploited for estimating the direction of the reaction plane~\cite{danielewicz_transverse_1985}.  For low event statistics, transverse Cartesian moments of the distributions can be evaluated.  Because of the uncertainty in the reaction plane determination, values of the moments, Cartesian or Fourier, need to be renormalized, on a moment by moment basis, and this may be done in a self-consistent manner \cite{danielewicz_transverse_1985, demoulins_measurement_1990, *demoulins_collective_1989, voloshin_flow_1996, poskanzer_methods_1998}.  However, for high event statistics, the whole distribution associated with the reaction plane could be addressed with the Fourier coefficients \cite{hades_collaboration_directed_2020}.  Taking another look, if such a distribution were evaluated directly using estimated directions for the reaction plane in collision events, the  distribution would come out blurred, because of the estimated direction of the plane fluctuates around the true direction from one event to another.
This brings up analogy to the situation in optics where a photo is taken with a camera that shakes.  For such situations and other, where distortions of image occur due to understood distortions of light intensity, deblurring techniques have been developed in the optical contexts~\cite{vankawala_survey_2015}.  
We propose to use such techniques in the context of central heavy-ion collisions to arrive at three-dimensional (3D) distributions tied to the reaction plane.  The interpretations of the latter distributions may be easier than of Fourier coefficients, especially those high, given that these coefficients tend to quantify information incidentally lumped together, through the reference to the same value of transverse momentum magnitude.  We moreover envision the use of such techniques to other situations in nuclear or high-energy physics where observables are distorted due to limitations of methodology or detector performance. 

We start out by reviewing the Richardson-Lucy (RL) deblurring algorithm \cite{richardson_bayesian-based_1972, lucy_iterative_1974}, that is popular in optics, based on Bayesian considerations, and well suited, in our opinion, for nuclear applications including heavy-ion collisions. We then turn to a schematic one-dimensional (1D) model, where a system moving at unknown velocity emits particles according to a single-particle distribution that is forward-backward symmetric in the c.m.\ frame.  A number of particles is measured and the goal is to determine the velocity distribution of particles relative to the center of mass, even though the center of mass is not known on the case by case basis.  The center of mass velocity may be estimated from the emitted particles, but it straggles relative to the true velocity.  With this, the distribution relative to the center of mass evaluated from the emitted particles alone is smeared out compared to the true distribution.  We demonstrate that the original distribution may be restored from the simulated observations by combining the central-limit theorem with the RL deblurring algorithm.  Next we repeat the model assuming that particles are measured with a detector for which the detection window is comparable with the spread of the particle distribution.  In this case, the central-limit theorem cannot be easily employed and the restoration problem becomes significantly nonlinear.  We demonstrate that in that more involved situation, the original velocity distribution may be still restored from the measured distribution following a self-consistent RL algorithm.  
We thereafter turn to simulated distributions for heavy-ion collisions.  We assume a typical situation recognized as a good representation of the final-state emission for semicentral collisions at few hundred MeV/nucl, with a local equilibrium combined with sideward, radial and elliptic flows.  With a moderate number of light charged particles registered, with mass $A \le 4$, it is possible to deblur their distributions associated with the reaction plane, even when these distributions vary by an order of magnitude or more in the transverse directions.  We complement the outline of the strategy for data analysis with results from transport theory, illustrating what kind of information could be accessed in the deblurred distributions, that might not be easily seen in the azimuthal moments for the distributions.

\section{Deblurring} \label{sec:deblur}

The blurring problem may be stated in the form of the equation:
\beq
n({\zeta}) = \int \text{d}{\xi} \, P({ \zeta} | { \xi}) \, {\mathcal N}({ \xi}) \, .
\label{eq:blur}
\eeq
Here, ${\mathcal N}({ \xi})$ is the distribution in coordinates ${ \xi}$ that faithfully characterizes the measured system and $P({ \zeta} | { \xi})$ is the probability density that the system is measured around $ \zeta$ when it is actually at $\bf \xi$.  Finally, $n({\bf \zeta})$ is the distribution attributed to the system in effect of the direct measurements.  The goal of deblurring is to determine ${\mathcal N}({ \xi})$ from the measured $n({\zeta})$, given knowledge about $P$.  While the problem appears stated as linear, in practical situations $P$ may have a dependence on ${\mathcal N}$, too.

In the reaction-plane problem, the blurring is due to the fact that the azimuthal angle relative to the estimated reaction plane is not equal to the angle relative to the true reaction plane.  The deviation may vary from one collision event to another and even particle to a particle, within the strategy of estimating the reaction plane orientation~\cite{danielewicz_transverse_1985, demoulins_measurement_1990}. In detector efficiency problems, $P$ might account for particle being misidentified or missed or whole event disregarded consequently under a trigger.  Depending on the situation, the probability density integrates to 1 or not, $\int_Z \text{d} \zeta \, P({ \zeta} | { \xi}) = P_Z(\xi) \le 1$.  For generality further on, we will consider the possibility of weighing the domain $Z$ with a function $W(\zeta)$, with the net weighted probability being then $P_{WZ}(\xi) = \int_Z \text{d} \zeta \, W(\zeta) \,  P({ \zeta} | { \xi})$.  The function $W$ may be then used to effectively shrink $Z$ such as in disregarding inferior quality events.

For inferring ${\mathcal N}$ directly from measured $n$, it might be tempting to invert \eqref{eq:blur} directly, but that strategy is likely to amplify short-wavelength noise always present in $n$.  In optical contexts alternative methods have been developed of which the Richardson-Lucy (RL) deconvolution algorithm \cite{richardson_bayesian-based_1972, lucy_iterative_1974}, based on Bayesian considerations that we next lay out, is particularly popular.

Let $Q(  { \xi} |  { \zeta})$ denote complementary probability density to $P$, that the system is at $\xi$ when it is measured at $\zeta$.  Then the probability that the system is within $\text{d}\xi$ while measured within $\text{d} \zeta$ can be expressed in two different ways:
\beq
Q(  { \xi} |  { \zeta}) \, n({\zeta}) \, \text{d} \zeta \, \text{d}  \xi = P({ \zeta} | { \xi}) \, {\mathcal N}({ \xi}) \, \text{d}  \xi \, \text{d} \zeta \, .
\eeq
This yields
\beq
Q(  { \xi} |  { \zeta}) = \frac{P({ \zeta} | { \xi}) \, {\mathcal N}({ \xi})}{ \int \text{d}{\xi'} \, P({ \zeta} | { \xi'}) \, {\mathcal N}({ \xi'})} \, ,
\eeq
and
\beq
{\mathcal N}({ \xi})  = \frac{\int \text{d}{\zeta} \, Q(  { \xi} |  { \zeta})\, W(\zeta) \, n({\zeta})}{P_{WZ} (\xi)} \, .
\label{eq:back}
\eeq
The RL method solves the last two equations iteratively:
\begin{eqnarray}
Q^{(r)}(  { \xi} |  { \zeta}) & = & \frac{P({ \zeta} | { \xi}) \, {\mathcal N}^{(r)}({ \xi})}{ \int \text{d} \xi' \, P({ \zeta} | { \xi'}) \, {\mathcal N}^{(r)}({ \xi'})} \, , \\[.5ex]
{\mathcal N}^{(r+1)}({ \xi})  & = & \frac{\int \text{d}{\zeta} \, Q^{(r)}(  { \xi} |  { \zeta}) \, W(\zeta) \, n({\zeta})}{P_{WZ} (\xi)} \, ,
\end{eqnarray}
where $r$ is iteration step index.
Any implicit dependence of $P$ on ${\mathcal N}$ may be handled in the equations iteratively.  One notable feature of the equations is that distributions that start as nonnegative stay such during the iterations.  Combination of the equations above yields
\beq
{\mathcal N}^{(r+1)}({ \xi})  = A^{(r)}(\xi) \cdot {\mathcal N}^{(r)}({ \xi})  \, ,
\eeq
where
\beq
A^{(r)} (\xi) = \frac{\int \text{d} \zeta  \,  \frac{n(\zeta)}{n^{(r)}(\zeta)}  \, W(\zeta) \, P(\zeta | \xi) } { \int \text{d} \zeta' \, W(\zeta') \, P(\zeta' | \xi)} \, ,
\eeq
and
\beq
n^{(r)}(\zeta) = \int \text{d}{\xi} \, P({ \zeta} | { \xi}) \, {\mathcal N}^{(r)}({ \xi}) \, .
\eeq
If the variable space is discretized, then the equations become
\beq
{\mathcal N}_i^{(r+1)} = A_i^{(r)} \cdot {\mathcal N}_i^{(r)} \, ,
\label{eq:iteration}
\eeq
\beq
A_i^{(r)} = \sum_j  \frac{n_j}{n_j^{(r)}  } \, T_{ji} \, ,
\label{eq:Ai}
\eeq
with
\beq
n_j^{(r)} = \sum_i P_{ji} \, {\mathcal N}_i^{(r)} \, ,
\label{eq:nN}
\eeq
and
\beq
T_{ji} = W_j \,  P_{ji} \Big/ \sum_{j'} W_{j'} \, P_{j'i} \, .
\eeq

To provide more discussion of the conventions and mathematics here, in the standard RL method, the variation of emphasis over the $Z$ domain is not employed, i.e., $W \equiv 1$.  We underscore the possibility of a different emphasis across~$Z$ for the sake of flexibility when the certainty in $n(\zeta)$ varies across the domain.  The variation in $W$ might also be used to test the robustness of conclusions.  Further on in this work we employ $W_j=1$, except for rare cases of $n_j^{(r)}=0$ when we put $W_j=0$.   Within the equations behind the RL method, Eq.~\eqref{eq:blur} represents forward mapping, while \eqref{eq:back} represents backward.  Depending on the context, the matrix $P_{ji}$ acting on a vector from the left in \eqref{eq:nN} may be called a blurring matrix or forward transfer matrix. The sum of elements in a column is bounded by~1, $\sum_j P_{ji} \le 1$, with the sum being exactly 1 when the probability is preserved.  The matrix $T_{ji}$ acting on a vector from the right in \eqref{eq:Ai} may be called a backward transfer matrix.  The sum of elements in a column of that matrix is exactly 1, $\sum_j T_{ji} = 1 $.

When $n$ and ${\mathcal N}$ extend over the same domain, it is common to use $n$ to start the iterations for ${\mathcal N}$.  It is common for the iterations to progress quickly first, but then gradually stall, with seesaw instabilities developing over many iterations.  Remedies include an acceleration \cite{remmele_vector_2009} by using a power $\nu > 1$ for the amplification factor $A$, terminating the iterations after a moderate number of steps, and/or using a regularization factor \cite{remmele_vector_2009, dey_richardsonlucy_2006}:
\beq
I^{(r)} = \frac{1}{1- \lambda {\bf H} \cdot {\bf \nabla} \, ({\bf \nabla} {\mathcal N}^{(r)} /|{\bf \nabla} {\mathcal N}^{(r)}|) } \,.
\eeq
Here, ${\bf H}$ is the discretization vector within the domain of ${\mathcal N}$ and $\lambda$ is a small factor that prevents build-up of any seesaw pattern in effect of the iterations.  I.e., Eq.~\eqref{eq:iteration} can be replaced with
\beq
{\mathcal N}_i^{(r+1)}({ \xi})  = \big[A_i^{(r)}(\xi)\big]^\nu \cdot I_i^{(r)} \cdot {\mathcal N}_i^{(r)}({ \xi})  \, ,
\label{eq:LRaccel}
\eeq
in an accelerated regularized LR scheme.  Excessive acceleration powers, $\nu > 2$, have been show to lead to an instability on its own in the iterations.

Obviously other methods of inferring $n$ from ${\mathcal N}$ are possible, such as in decomposing them in orthogonal functions, including harmonics \cite{voloshin_flow_1996, hades_collaboration_directed_2020}.  The danger in the latter case is of combining functions with varying sign where noise is handled independently.  In the case of functions that change by orders of magnitude, the danger is of arriving at a poorer representation for ${\mathcal N}$ in the regions of low values than even provided by $n$.

\section{1D Model}

We next present a schematic and thus potentially transparent 1D model.  It may at first seem disconnected from the issue of reaction-plane deblurring.  In reality, it is closely related.

We consider a projectile moving at an unknown velocity $V$ that varies from an event to an event.  The projectile de-excites emitting $N=10$ identical particles sampled from a uniform velocity distribution from -1 to +1 relative to $V$, in some arbitrary velocity units.  The particle velocities are measured and the goal is to determine, from event statistics, the distribution of particles in the projectile frame.  For uncorrelated emission, the rms of the sought distribution is obviously easy to estimate from the average difference of velocities between any two particles squared:
\beq
\langle (v_1^\text{lab} - v_2^\text{lab})^2 \rangle = \langle (v_1 - v_2)^2 \rangle = 2 \langle v^2 \rangle \, .
\label{eq:1Dsquare}
\eeq
Here, $v$ are velocities relative to the projectile and $v^\text{lab}$ are in the laboratory system: $v^\text{lab} = v + V$.

\begin{figure}
\centering
\includegraphics[width=.6\linewidth]{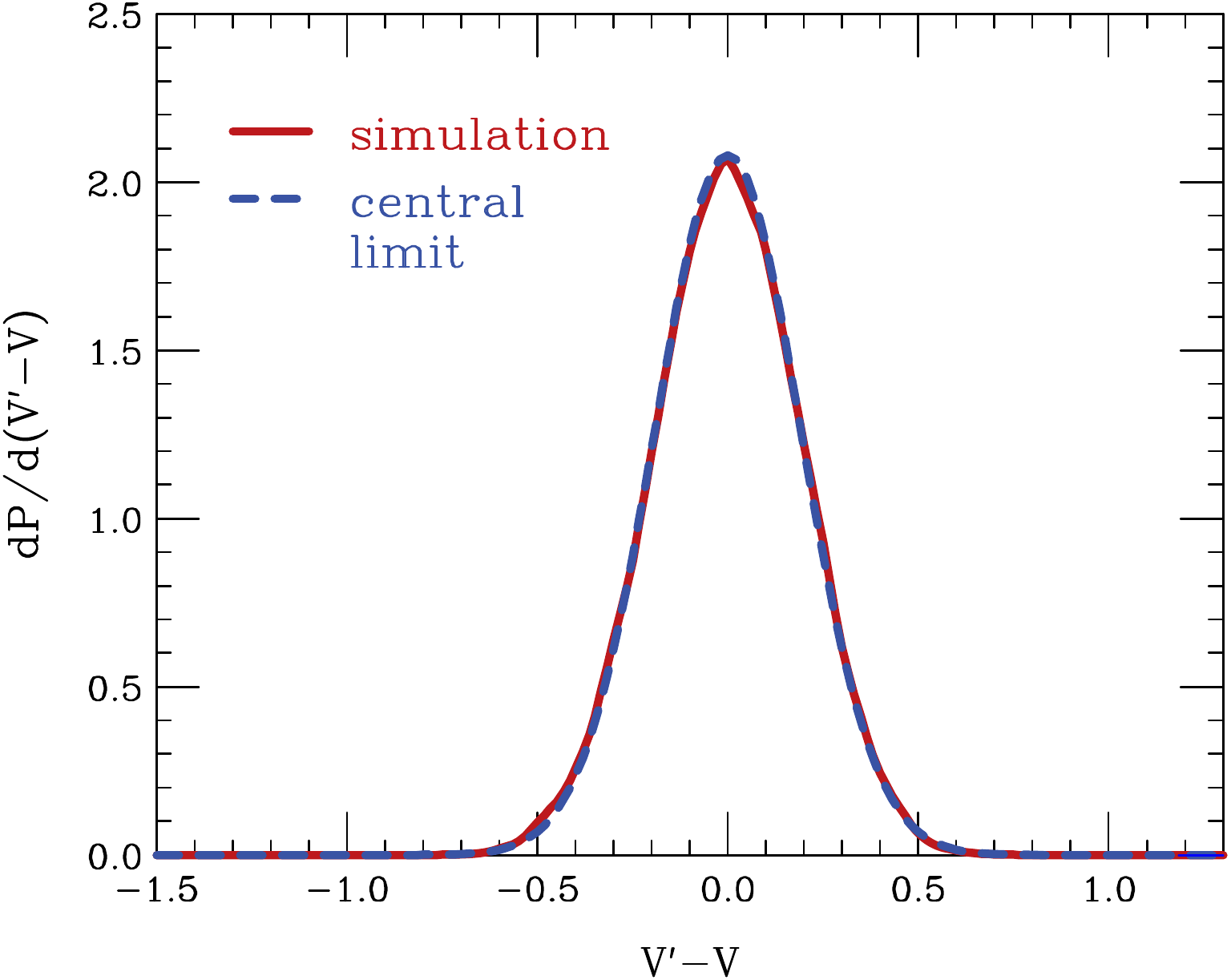}
\caption{Probability density for system velocity $V'$, estimated with the average velocity for $N-1=9$ particles sampled from a symmetric uniform velocity distribution centered around the true velocity $V$.  The solid (red) line shows the distribution from a direct simulation, calculated with $H=0.1$ binning, and the dashed (blue) line - the distribution estimated with the central-limit theorem.}
\label{fig:1D}
\end{figure}

In assessing velocity of a particle relative to the projectile, the velocity of the projectile can be estimated with the average velocity of  $N-1 = 9$ remaining particles, in a similar manner as when reaction plane is estimated from remaining particle in an event \cite{danielewicz_transverse_1985}.  The so assessed velocity $V'$ fluctuates though around the true velocity $V$.  The probability distribution from  simulated 10,000 events is shown as a solid curve in Fig.~\ref{fig:1D}.  The central-limit theorem states that for a large particle number in one event such a distribution approaches a Gaussian
\beq
\frac{\text{d} P}{\text{d} V'} \simeq \sqrt{\frac{(N-1) }{2 \pi \langle v^2 \rangle}} \, \text{exp} \bigg[ - \frac{(N-1)(V' - V)^2}{2  \langle v^2 \rangle} \bigg] \, .
\label{eq:1DGaussian}
\eeq
We compare that Gaussian distribution to the one from simulations and it is apparent that the two distributions are nearly indistinguishable in the region where they are significant.

\begin{figure}
\centering
\includegraphics[width=.55\linewidth]{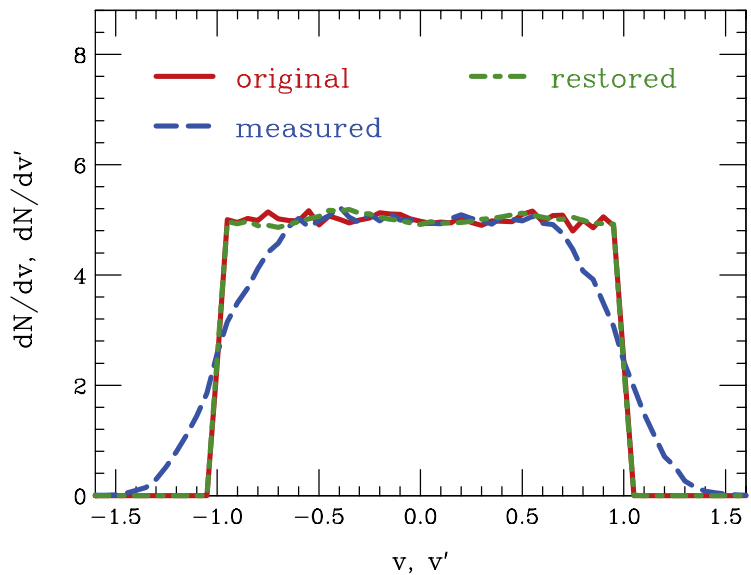}
\caption{Distribution of particles in velocity within the projectile frame for 1D model: original represented by the solid (red) line, inferred from measurements represented by the dashed (blue) line and restored through deblurring and represented by the dash-dot (green) line.  In the simulation, 10,000 of events with $N=10$ particles each were generated.  A binning of $H=0.1$ was used for the distributions.  The restoration was done assuming the central-limit blurring function.}
\label{fig:1Dx}
\end{figure}

Because of the fluctuations of the reference $V'$, around the true velocity $V$, the inferred velocities $v'$ relative to the projectile are inaccurate, $v'$ = $v + V - V'$, and the inferred distribution in velocity gets smeared out:
\beq
\frac{\text{d} N}{\text{d} v'}(v') = \int \text{d}V' \, \frac{\text{d}P}{\text{d}V'}(V' - V) \, \frac{\text{d} N}{\text{d} v}(v) \, .
\eeq
In Fig.~\ref{fig:1Dx}, we show the original distribution of particles relative to the projectile in the simulations, together with that from simulated measurements and the smearing is evident.  In collecting the statistics, we use $H=0.1$ bin size.  The third distribution represented in Fig.~\ref{fig:1Dx} is one from deblurring using the regularized and accelerated RL algorithm.  In 1D, the regularization factor is
\beq
I_i^{(r)} = \begin{cases}
              \frac{1}{1-\lambda} \, , & \mbox{if } \, \, {\mathcal N}_i^{(r)} < {\mathcal N}_{i-1,i+1}^{(r)} \, , \\
              \frac{1}{1+\lambda}\, , & \mbox{if } \, \,   {\mathcal N}_i^{(r)} > {\mathcal N}_{i-1,i+1}^{(r)} \, , \\
              1 \, , & \mbox{otherwise} \, .
            \end{cases}
\eeq
We employ $\lambda = 0.01$ and accelerating power $\nu=1.99$ and start with ${\mathcal N}^{(1)} = n$.  Moreover, we carry out the iterations in \eqref{eq:LRaccel} by using the central-limit blurring function \eqref{eq:1DGaussian} with \eqref{eq:1Dsquare}.  In principle, it could be possible to update the blurring functions during iterations to make it consistent with the inferred d$N$/d$v$, for finite $N$, if desired accuracy of restoration called for that.  At the level of the statistics we employ, the iterations that employ our more basic procedure stabilize at a distribution that is largely consistent with the original.

\section{Model with Detection Inefficiencies}

Here, we expand the 1D model of the previous section assuming detection inefficiencies.  This continuous a connection to the reaction-plane deblurring, though maybe the particular challenges are overemphasized compared to a typical case of the reaction-plane determination.  The overemphasis tests the robustness of the restoration procedures which can be useful for other applications in nuclear physics.

\begin{figure}
\centering
\includegraphics[width=.60\linewidth]{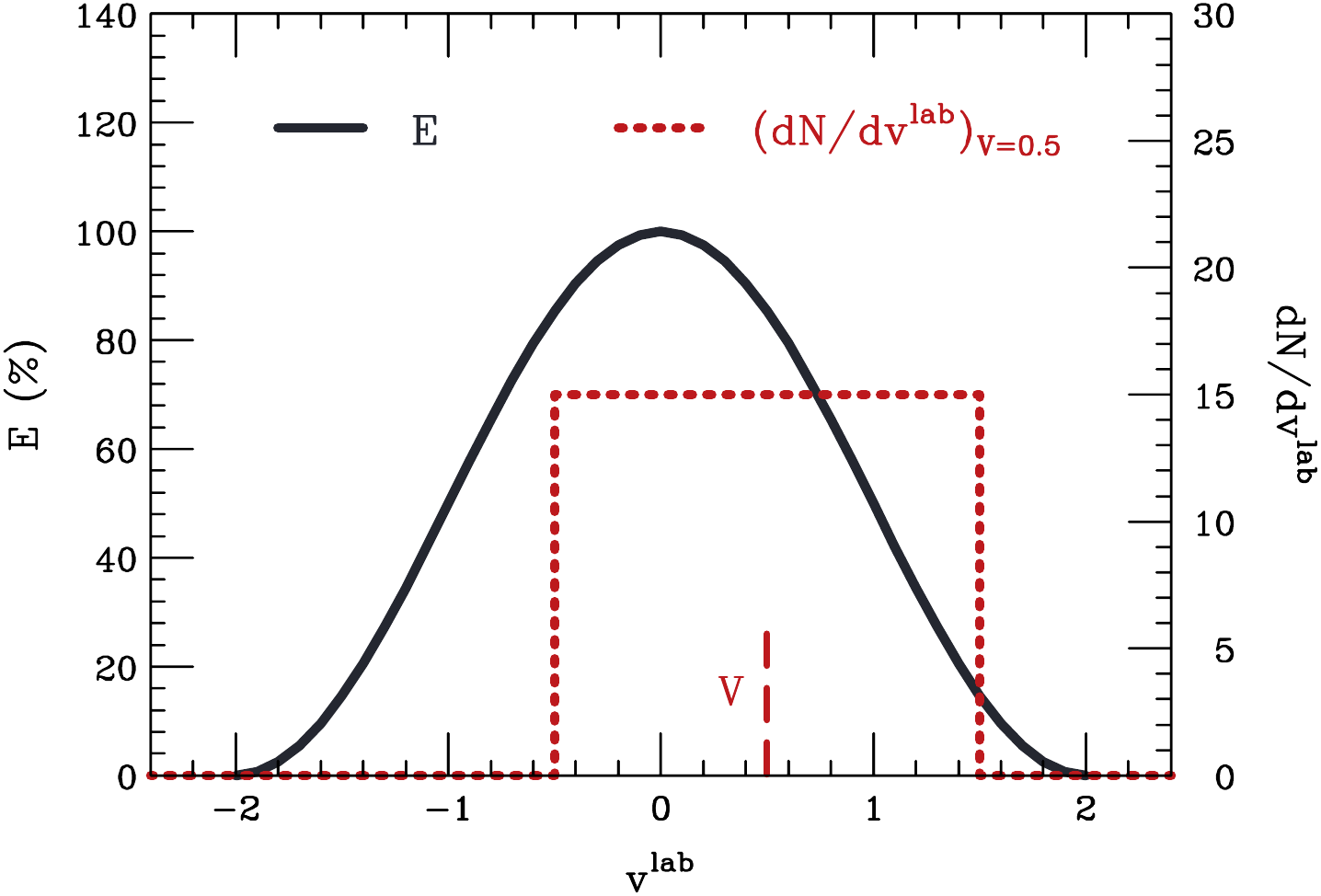}
\caption{Detection efficiency $E$ (solid line) as a function of particle laboratory velocity $v^\text{lab}$, in the 1D model with inefficiencies, with superimposed particle distribution $dN/dv^\text{lab}$ (dotted line) for a projectile moving at velocity $V=0.5$ (short vertical dashed line).}
\label{fig:detection}
\end{figure}

Again in the model we assume that a projectile moves at unknown velocity $V$ that varies from an event to event.  The projectile de-excites emitting $N$ identical particles sampled from a uniform velocity distribution from -1 to +1 relative to $V$, in the arbitrary velocity units. The particle velocities are measured and the goal is to determine, from event statistics, the distribution of particles in the projectile frame.  Now, however, we assume that the particles are detected using a detector with efficiency $E(v^\text{lab})$ varying across the detection window  $|v^\text{lab}| < v^\text{lab}_\text{max} = 2$:
\beq
E(v^\text{lab}) = \begin{cases}
      \frac{1}{2} + \frac{1}{2} \cos{\Big(\pi \, v^\text{lab} /v^\text{lab}_\text{max} \Big)} \, ,  & \mbox{if} \, \,  |v^\text{lab}| <  v^\text{lab}_\text{max} \, , \\  0 \, , & \mbox{otherwise} \, . \\\end{cases}
\eeq
The efficiency is illustrated in Fig.~\ref{fig:detection} with the superimposed distribution of emitted particles for an exemplary projectile velocity $V=0.5$.  Given that the number $n$ of detected particles is generally reduced compared to the number $N$ of emitted, $n \le N$, we assume now $N=30$ particles and demand that a minimum $n_\text{min} = 10$, $n \ge n_\text{min} $ is detected for the events to be analysed with the goal of finding $\text{d}N/\text{d}v$ in the projectile frame. 

\begin{figure}
\centering
\includegraphics[width=.60\linewidth]{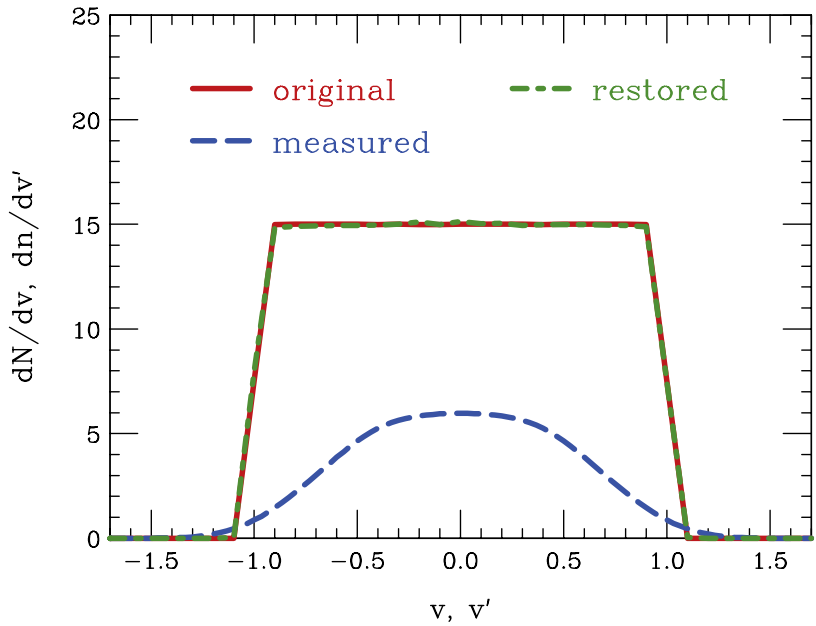}
\caption{Distribution of particles in velocity within the projectile frame for 1D model with inefficiencies: original represented by the solid (red) line, inferred from measurements represented by the dashed (blue) line and restored through deblurring and represented by the dash-dot (green) line.  The simulated events originally contained $N=30$ particles.  In the measurement the particles were assumed to be accepted with the efficiency of Fig.~\ref{fig:detection} and the events were processed if the number of accepted particles exceeded a minimum: $n \ge n_\text{min} = 10$.  A binning of $H=0.1$ was used for the distributions and in the restoration matrices.}
\label{fig:1dex}
\end{figure}

In the sampling of the events, we distribute the projectile velocity $V$ uniformly in the laboratory frame over the range $|V | < V_\text{max} = v_\text{max} + v^\text{lab}_\text{max} + H/2$.  Here, $H=0.1$ continues to be our bin size employed in discretizing distributions.  The precise value of $V_\text{max}$ is ultimately of a minor importance, as the probability of accepting an event for analysis becomes very small when $|V|$ gets close to $v^\text{lab}_\text{max}$ terminating the detector acceptance.  Varying $V_\text{max}$ in some range above then largely amounts to adding or subtracting events that do not get analysed.  Average distribution of particles relative to the projectile velocity, for the original events and for the simulated measurements, is illustrated in Fig.~\ref{fig:1dex}.  Note that the average in Fig.~\ref{fig:1dex} is over the simulated events, that may or may not be analysed.  Any extra events that do not get analysed, due to $n < n_\text{min}$, contribute a drop in the overall normalization of $\text{d}n/\text{d}v'$ distribution found in analysing the events, without altering the shape of that distribution. That normalization eventually drops out in the restoration.

In determining a contribution to $\text{d}n/\text{d}v'$ from each analysed event, we progress as before and for any particular particle we estimate the projectile velocity $V'$ as the average velocity of the remaining $n-1$ particles in an event.  In spite of the starting particle number being significantly higher than before and only the low cut-off for observed particle number $n_\text{min}$ coinciding with the prior starting number, the shape of the measured distribution in Fig.~\ref{fig:1dex} is now much more degraded compared to the original, than in the previous model consideration of Fig.~\ref{fig:1Dx}.  The impact of the significant, rapidly changing inefficiency on the measurement, that can be expressed with conditional probabilities, and on restoration, is discussed below.

\begin{figure}
\centering
\includegraphics[width=.57\linewidth]{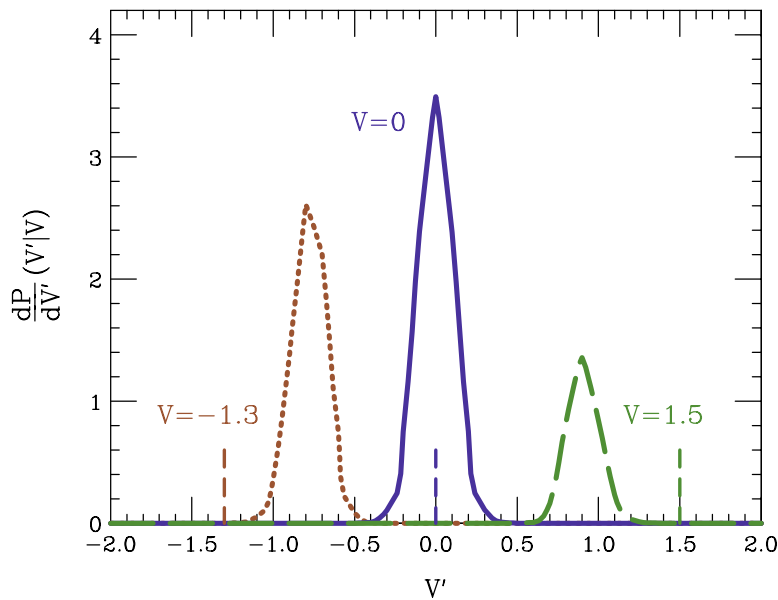}
\caption{Conditional probability density for determining velocity $V'$, from decay products in the 1D model with inefficiencies, at three indicated velocities $V$ of the projectile.  Those original velocities $V$ are marked with the short vertical dashed lines by the abscissa.  Following the strategy of using the remainder of the system as a reference for any single particle, in generating the illustrated density, $N-1 = 29$ particles can be sampled for an incomplete event and the set analysed if at least $n_\text{min} -1 =9$ particles pass the detector acceptance.}
\label{fig:1DE}
\end{figure}

On the technical side, the probability densities for determining the system velocity to be $V'$, while the original is~$V$, and for the particle velocity to be $v'$ in the system frame, when the  original is $v$, do not depend anymore on just difference of density arguments, as in previous example, but on each argument separately, and these densities are not identical to each other anymore.  In the measurement, when the projectile velocity $V$ pushes against a region of reduced detection efficiency, one can expect the velocity $V'$, determined from decay products, to be biased towards the region of higher efficiency.  This is indeed seen in terms of conditional probability density $\frac{\text{d}P}{\text{d}V'}(V'|V)$, the examples of which are shown in Fig.~\ref{fig:1DE}.  Also expected drop in the overall probability of an event getting accepted for the analysis can be seen as~$V$ gets closer to the particle detection limits.  As far as the width of the distributions for a given $V$ is concerned, the quickly changing inefficiency has a dual effect, of increasing the width due to a reduced accepted particle number and decreasing due to shrinking of the accepted particle distribution.  In the end, the width is only modestly changing with $V$ and it is actually shrunk compared to the previous case with 100\% detection but fewer particles at the start.

\begin{figure}
\centering
\includegraphics[width=.57\linewidth]{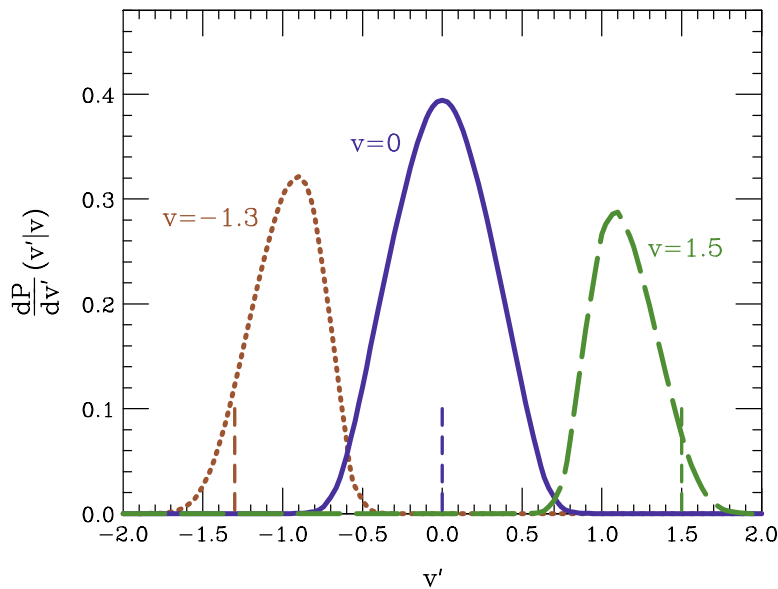}
\caption{Conditional probability density for finding velocity $v'$ relative to the projectile, for a particle measured in an event in the 1D model with inefficiencies, at three indicated true velocities $v$ relative to the projectile.  Those original velocities $v$ are marked with the short vertical dashed lines by the abscissa.}
\label{fig:1Dde}
\end{figure}

The conditional probability for the velocity relative to projectile frame, $\frac{\text{d}P}{\text{d}v'}(v'|v)$, can be represented as a convolution of the efficiency function $E$ with the conditional probability for the projectile velocity, $\frac{\text{d}P}{\text{d}V'}(V'|V)$:
\beq
\frac{\text{d}P}{\text{d}v'}(v'|v) = \frac{1}{2V_\text{max}} \int_{-V_\text{max}}^{V_\text{max}} \text{d}V \, E(V+v) \, \frac{\text{d}P}{\text{d}V'}(V+ v-v'|V) \, .
\label{eq:dPv}
\eeq
The conditional probability density at exemplary values of the original velocity $v$, from the model simulations, is illustrated in Fig.~\ref{fig:1Dde}.  To underscore, that density is established no matter what the value of $\text{d}N/\text{d}v$ at $v$ is, even when that value is zero.  Though then that last distribution might seem to disappear from the ongoing consideration, it is obviously there in $\frac{\text{d}P}{\text{d}v'}(v'|v)$, as the estimation of the projectile velocity $V'$ depends on the overall $\text{d}N/\text{d}v$, within $\frac{\text{d}P}{\text{d}V'}(V'|V)$ in~\eqref{eq:dPv}.  In the model in the previous section, that inherent dependence could be largely described in terms of the easily estimated dispersion for the distribution, but here this dependence is generally more complex.  The measured distribution of particles in relative velocity, $\text{d}n/\text{d}v'$, is related to the original distribution $\text{d}N/\text{d}v$ with, cf.~Eq.~\eqref{eq:blur},
\beq
\frac{\text{d}n}{\text{d}v'} = \int \text{d}v \, \frac{\text{d}P}{\text{d}v'}(v'|v) \, \frac{\text{d}N}{\text{d}v} \, .
\label{eq:dnN}
\eeq

In comparing the conditional probability density for relative velocity $\frac{\text{d}P}{\text{d}v'}(v'|v)$, of Fig.~\ref{fig:1Dde} to the density for projectile velocity $\frac{\text{d}P}{\text{d}V'}(V'|V)$, of Fig.~\ref{fig:1DE}, we can see that the density for the relative velocity is much wider, exhibits nearly as much bias as the other for finite velocity values, and it more clearly exhibits skewness.  
The width and bias in the conditional $\frac{\text{d}P}{\text{d}v'}(v'|v)$ have their origin in the widths and biases of contributing conditional $\frac{\text{d}P}{\text{d}V'}(V'|V)$ in Eq.~\eqref{eq:dPv}.  The strong increase in the widths for $\frac{\text{d}P}{\text{d}v'}(v'|v)$ can be attributed to the range of  $\frac{\text{d}P}{\text{d}V'}(V'|V)$ contributions, with different~$V$ and changing bias.  
Regarding the bias, it may be observed in Eq.~\eqref{eq:dPv} and otherwise, that $v'-v = -(V -V')$, so the bias contributed to $\frac{\text{d}P}{\text{d}v'}(v'|v)$ by events is opposite to the bias contributed to  
$\frac{\text{d}P}{\text{d}V'}(V'|V)$.  The fact that the bias is primarily negative in the forward direction for $\frac{\text{d}P}{\text{d}v'}(v'|v)$ in Fig.~\ref{fig:1Dde}, means that the forward relative velocity will be primarily measured for events where the projectile velocity is in the negative direction, and the other way around.  This is consistent with the singled out velocity and reference particle set fitting in an optimal manner within the detector acceptance window, one pushing the front and the other the rear of that window.  The skewness can be tied to the fact that large widths become comparable to the pace at which efficiency is changing.  
Both the probability for analysing an instance of a given relative velocity $v$, $P(v) = \int \text{d}v' \, \frac{\text{d}P}{\text{d} v'}(v'|v)$, and net velocity~$V$, $P(V) = \int \text{d}V' \, \frac{\text{d}P}{\text{d} V'}(V'|V)$, fall with increase in magnitude of the respective velocity, see Figs.~\ref{fig:1DE} and~\ref{fig:1Dde}.
Overall, the large width of the conditional probability density  $\frac{\text{d}P}{\text{d}v'}(v'|v)$, larger for central $v$ than in the model of the preceding section, the bias, not there before in the model, and the fall of the analysis probability with $|v|$, again not there before, all contribute to much stronger deterioration of the shape of distribution in relative velocity in the measurements compared to the original shape, cf.~Figs.~\ref{fig:1Dx} and \ref{fig:1dex}.

For restoration of $\text{d}N/\text{d}v$, Eq.~\eqref{eq:dnN} is discretized.  The difficulty now is that the conditional probability density at iteration $r$, $\frac{\text{d}P^{(r)}}{\text{d}v'}(v'|v)$, depends in a nontrivial manner on ${\text{d}N^{(r)}}/{\text{d}v}$.  We start the iterations of Sec.~\ref{sec:deblur} with ${\text{d}N^{(1)}}/{\text{d}v} = \langle E \rangle^{-1} \, {\text{d}n}/{\text{d}v'}$, at $v'=v$, and we compute $\frac{\text{d}P^{(1)}}{\text{d}v'}(v'|v)$ through a simulation.  Here, $\langle E \rangle$ is the typical detector acceptance that we conservatively take as average efficiency over the detection window, i.e., $\langle E \rangle = 0.5$.  Simulations for determining $\frac{\text{d}P^{(r)}}{\text{d}v'}(v'|v)$ are numerically costly compared to the matrix multiplications in the RL algorithm and this is likely to be the case in similar situations of restoration.  We tame increase in the cost by updating $\frac{\text{d}P^{(r)}}{\text{d}v'}(v'|v)$ only once in a while during the iterations, infrequently enough for the RL iterations on their own to converge in-between.  As is illustrated in Fig.~\ref{fig:1dex}, despite of the strong detector inefficiencies, strongly distorted measured distribution, and greater complexity in the restoration, the procedure can restore the original single-particle distribution sufficiently well for typical practical applications.

\section{Reaction-Plane Deblurring}

The problem of the blurring of particle distributions and other characteristics of heavy-ion collisions lies in attributing azimuthal angle $\Phi'$ to the reaction plane that is actually directed at angle $\Phi$.  With this, the angles relative to the reaction plane attributed to the particles are inaccurate, $\phi'=\phi+\Phi-\Phi'$, and measured distributions are going to be smeared out:
\beq
\frac{\text{d}^3 \, N}{p_\perp \, \text{d}p_\perp \, \text{d} y \, \text{d}\phi' } = \int \text{d} \Phi' \, \frac{\text{d}P}{\text{d} \Phi'}(\Phi'-\Phi) \,
\frac{\text{d}^3 \, N}{p_\perp \, \text{d}p_\perp \, \text{d} y \, \text{d}\phi } (\phi) \, .
\label{eq:d3N}
\eeq

Owing to the asymmetries in emission tied with the reaction plane, the direction of the reaction plane is estimated with a combination of particle momenta, most commonly sum of weighted transverse momenta, ${\bf q}_\nu = \omega_\nu \, {\bf p}_{\perp \nu}$.  When the direction of the reaction plane is used as a reference for a certain particle $\mu$, that particular particle is omitted from the reference, to avoid a self-correlation.  I.e., the reaction plane direction is estimated with the direction of the vector \cite{danielewicz_transverse_1985, andronic_systematics_2006}
\beq
{\bf Q}_{\mu'} = \sum_{\nu \ne \mu } \omega_\nu \, {\bf p}_{\perp \nu} = \sum_{\nu \ne \mu} {\bf q}_\nu \, .
\label{eq:Q}
\eeq
The direction is obviously the same as for the average vector ${\bf q}$ of the remaining particles
\beq
\overline{\bf q_{\mu'}} = \frac{1}{N-1} \sum_{\nu \ne \mu} {\bf q}_\nu \, ,
\label{eq:qmu}
\eeq
where $N$ is particle multiplicity.  When ${\bf q}_\nu$ are largely uncorrelated, except through the reaction-plane geometry, the central-limit theorem implies
the distribution of the average vector $\overline{\bf q}$ in relation to the reaction plane approaches Gaussian for large $N$:
\beq
\frac{\text{d}^2 P}{\text{d}^2 \overline{q}} \simeq \frac{N-1}{2 \pi \, \sigma_x \, \sigma_y} \, \text{exp} \bigg\lbrace -\frac{N-1}{2} \bigg[ \frac{\big(\overline{q}_x - \langle q_x \rangle \big)^2}{\sigma_x^2} +  \frac{\overline{q}_y^2}{\sigma_y^2} \bigg]\bigg\rbrace \, .
\label{eq:d2P}
\eeq
Here, index $x$ is for a component along the reaction-plane direction and $y$ -- along the perpendicular direction.  The dispersions $\sigma$ and average $\langle q_x \rangle$ refer to single-particle emission in relation to the reaction plane,
\beq
\label{eq:sigmas}
\begin{split}
\sigma_x^2 & = \langle (q_x - \langle q_x \rangle)^2 \rangle = \langle q_x^2 \rangle - \langle q_x \rangle^2 \, , \\
\sigma_y^2 & = \langle q_y^2 \rangle \, ,
\end{split}
\eeq
Under the condition that the correlations tied to the reaction plane dominate \cite{danielewicz_collective_1988}, the parameters for the probability distribution can be determined from constructs out of the ${\bf q}$ contributions averaged over particle sets and events:
\beq
\label{eq:products}
\begin{split}
  \langle {\bf q}_1 \cdot {\bf q}_2 \rangle & = \langle q_x \rangle^2 \, , \\
 \langle 2({\bf q}_1 \cdot {\bf q}_2)({\bf q}_2 \cdot {\bf q}_3) - ({\bf q}_1 \cdot {\bf q}_3) \, q_2^2\rangle & = \langle q_x \rangle^2 \, (\langle q_x^2 \rangle - \langle q_y^2 \rangle ) \, .
\end{split}
\eeq
Here, the indices 1, 2 and 3 refer to any three particles contributing to ${\bf Q}$ in an event.

Now, we can express $\overline{\bf q}$ in terms of azimuthal deviation from the true direction of the reaction plane, $\Delta \Phi = \Phi' - \Phi$
\beq
\overline{q}_x = \overline{q} \cos{\Delta \Phi} \, , \hspace*{2em} \overline{q}_y = \overline{q} \sin{\Delta \Phi} \, ,
\eeq
and write the probability density \eqref{eq:d2P} as
\beq
\frac{\text{d}^2 P}{\text{d}^2 \overline{q}} = \frac{\text{d}^2 P}{\overline{q} \, \text{d}\overline{q} \, \text{d} \Phi'} \simeq {\mathcal A} \, \text{exp} \big[-{\mathcal B}(\Delta \Phi)\, \overline{q}^2+2 {\mathcal C} (\Delta \Phi)\, \overline{q} \big] \, .
\eeq
Integration over magnitude of $\overline{q}$ yields the central-limit distribution of the estimated reaction plane relative to the true plane:
\beq
\frac{\text{d}P}{\text{d}\Phi'} = \int \text{d} \overline{q} \,  \overline{q} \, \frac{\text{d}^2 P}{\text{d}^2 \overline{q}}
\simeq \alpha (\Delta \Phi) \, \Big\lbrace 1 + \sqrt{\pi} \, \beta(\Delta \Phi) \, \exp{\big(\beta^2(\Delta \Phi)\big)} \big[ 1 + \text{erf} \big( \beta (\Delta \Phi) \big)  \big] \Big\rbrace \, .
\label{eq:dPdPhi}
\eeq
Here, the terms $\beta$ and $\alpha$, both dependent on the deviation $\Delta \Phi$ from the true reaction plane, are
\beq
\beta (\Delta \Phi) = \frac{{\mathcal C} (\Delta \Phi)}{\sqrt{{\mathcal B}(\Delta \Phi)}} = \sqrt{\frac{N-1}{2\big(\sigma_y^2 \cos^2{\Delta \Phi} + \sigma_x^2 \sin^2{\Delta \Phi} \big)}} \, \frac{\sigma_y}{\sigma_x} \, \langle q_x \rangle \cos{\Delta \Phi} \approx \sqrt{\frac{N-1}{\langle q^2 \rangle}} \, \langle q_x \rangle \, \cos{\Delta \Phi} \, .
\label{eq:beta}
\eeq
and
\beq
\alpha (\Delta \Phi) = \frac{\mathcal A}{2 {\mathcal B}(\Delta \Phi)} = \frac{\sigma_x \, \sigma_y \, \exp{\big[-\frac{(N-1)\langle q_x \rangle^2}{2 \sigma_x^2}\big]}}{2\pi \, \big( \sigma_y^2 \, \cos^2{\Delta \Phi} + \sigma_x^2 \, \sin^2{\Delta \Phi}\big)} \approx \frac{1}{2\pi} \, \exp{\bigg[-\frac{(N-1)\langle q_x \rangle^2}{\langle q^2 \rangle}\bigg]}  \, ,
\label{eq:alpha}
\eeq
see also Ref.~\cite{voloshin_flow_1996}.
The final approximations in Eqs.~\eqref{eq:beta} and \eqref{eq:alpha} pertain to the case of weak azimuthal anisotropies in the distributions relative to the reaction plane.  These underscore that the main variation with angle in \eqref{eq:dPdPhi} is tied to the factor of $\cos{\Delta \Phi}$ in \eqref{eq:beta}.

An example of the distribution \eqref{eq:dPdPhi} is shown in Fig.~\ref{fig:dPhi}.  We next illustrate the envisioned procedure of reaction plane deblurring in the case of the final state modelled using a local-equilibrium model that incorporates common concepts for intemediate-energy heavy-ion collisions.

\subsection{Deblurring Example: Local-Equilibrium Model}

We illustrate deblurring for a 3D distribution associated with the reaction plane, deducing $\text{d}^3 \, N/p_\perp \, \text{d}p_\perp \, \text{d} y \, \text{d}\phi$ from $\text{d}^3 \, N/p_\perp \, \text{d}p_\perp \, \text{d} y \, \text{d}\phi'$, c.f.\ Eq.~\eqref{eq:d3N}, by simulating the final state of a collision in a local equilibrium model.  As a reference beam energy for the collision, we take $300 \, \text{MeV/nucl}$ and consider $A \le 4$ charged products in the final state.

\begin{figure}
\centering
\includegraphics[width=.55\linewidth]{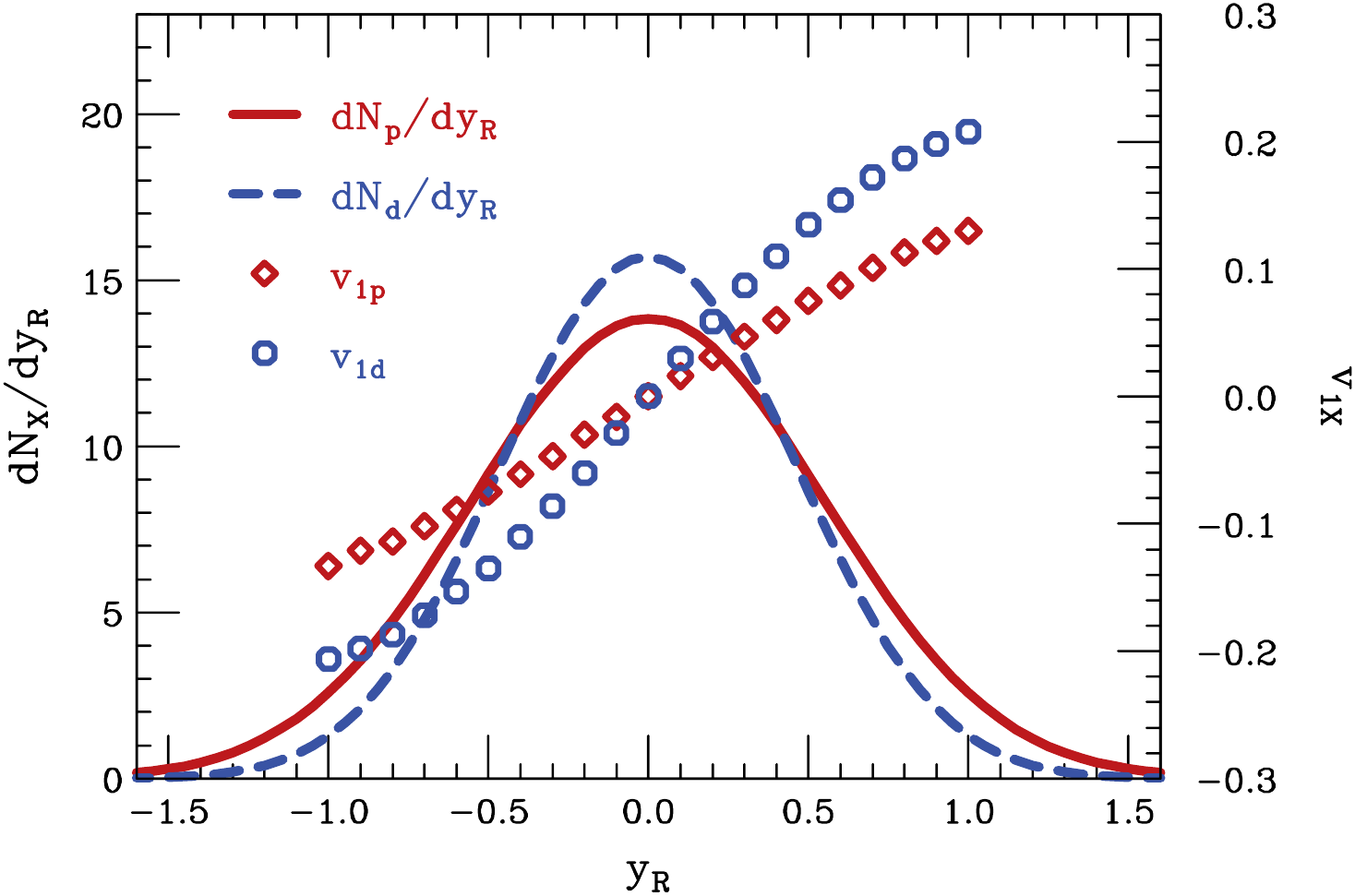}
\caption{Characteristics of protons and deuterons within the local equilibrium model.  The lines represent differential distributions in normalized rapidity, $\text{d}N_X/\text{d}y_R$, and symbols represent first-order azimuthal-asymmetry coefficients, $v_{1X} = \langle \cos{(\phi-\Phi)} \rangle_X$, all vs $y_R$.  Here, the rapidity is in the center of mass and scaled with the rapidity of the beam, $y_R=y/y_\text{Beam}$.}
\label{fig:dNv1}
\end{figure}

Our goal is  to mimic a state that is consistent at a very coarse level with the observations \cite{andronic_systematics_2006, reisdorf_systematics_2010}.  We have no intention to dive into any controversies around the freeze-out in collisions,  whether the colliding systems literarily go through an equilibrium when expanding into vacuum or whether they separately freeze out chemically and kinematically.  Specifically, we assume that the system freeze-out can be described in terms of a local equilibrium, at a uniform freeze-out density $\rho_f$ and temperature $T_f$, combined with collective motion for the freeze-out location.   Within the latter, the longitudinal expansion dominates, but also present are a radial transverse expansion, with a weak elliptical modulation, and finally a weak sideward flow, typical for mid-central collisions.  We take the freeze-out density at $\rho_f = \rho_0/6$ and, for simplicity, a neutron-proton symmetric system.  Assuming that the local kinetic energy for nucleons is a fraction of the kinetic energy available per nucleon, we arrive at the freeze-out temperature $T_f \simeq 29 \, \text{MeV}$.  To fix the attention, we assume that the average number of protons included in the measurements, whether as free or in clusters, is $\langle Z_t \rangle = 50$.  Then, $\rho_f$ and $T_f$ give us charged particle multiplicities, $\langle N_p \rangle= 19.0$, $\langle N_d \rangle  = 17.8$, $\langle N_t \rangle = \langle N_h \rangle =3.7$, and $\langle N_\alpha \rangle = 1.0$.  The $A \ge 3$ multiplicities are low, for a given proton multiplicity, compared to the data from the specific beam energy region \cite{reisdorf_systematics_2010}.  However, our main goal is to explore the situation with different species used in reaction plane determination, not a literal reproduction of relative yield data.  We sample multiplicities for the species from Poisson distributions for the average specie multiplicities and we combine a momentum sampled from a local equilibrium distribution for the specific species with a collective velocity boost that itself combines the aforementioned persistence of longitudinal motion, radial transverse expansion and elliptic and sideward flows, all of the latter common between the species.  In Fig.~\ref{fig:dNv1}, we show exemplary characteristics of protons and deuterons in the generated events.  Second-order coefficients of azimuthal anisotropy have low values in this simulation, in fact for protons quite comparable to statistical errors even for a large number of events, and are not shown.

\begin{figure}
\centering
\includegraphics[width=.55\linewidth]{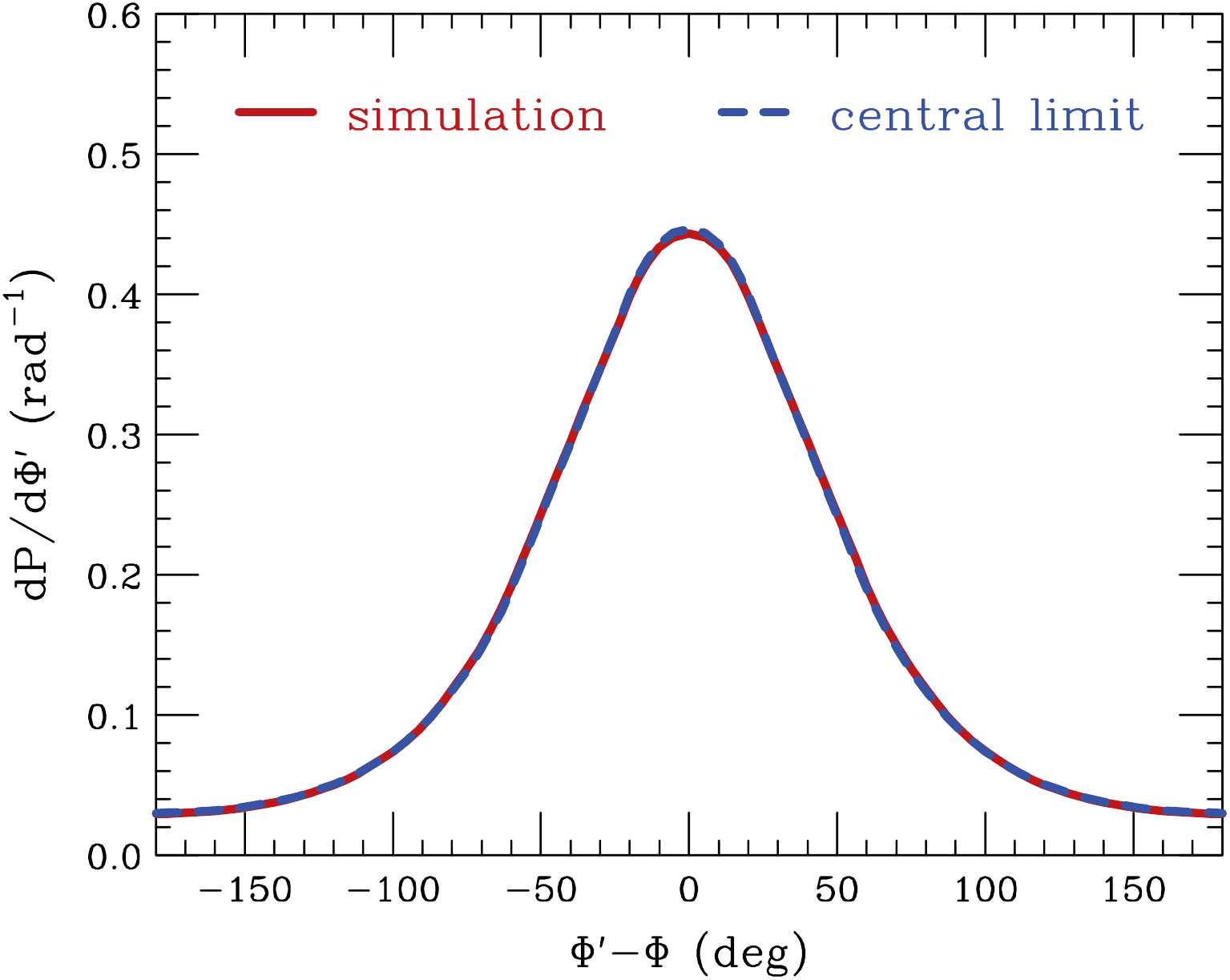}
\caption{Probability density for the azimuthal angle $\Phi'$ of the reaction plane in the local equilibrium model, when estimated with the direction of the vector $\overline{\bf q}$ out of $N - 1$ particles in an event, relative to the true angle~$\Phi$.  Here, $N$ is the total particle number in an event.  The solid (red) line shows the density from a direct simulation in the local equilibrium model and the dashed (blue) line shows the density estimated with the central-limit Eq.~\eqref{eq:dPdPhi}.  In using the last equation, $\langle q_x \rangle$, $\sigma_x^2$ and $\sigma_y^2$ in $\alpha$ and $\beta$ were obtained from averages over the event simulations, following the strategy laid out in the text.  In addition, the multiplicity $N$, actually varying from an event to event, was replaced with the single average value $\langle N \rangle$ over the simulated events.}
\label{fig:dPhi}
\end{figure}

In each of the events in the sample, when considering particle $\mu$, we estimate the reaction plane direction with the remaining particles in the event.  We use
\beq
\omega_\nu =
\begin{cases}
\text{sgn} \, y_R \, , & \text{if $|y_R| > \delta$}, \\
0 \, , & \text{if $|y_R| < \delta$},
\end{cases}
\label{eq:omega}
\eeq
in Eq.~\eqref{eq:Q}, with $\delta=0.17$ \cite{danielewicz_transverse_1985}.  The distribution of estimated plane directions, relative to the true direction, is shown in Fig.~\ref{fig:dPhi} for this choice of weights.  More optimal weights could be chosen, yielding a more narrow distribution \cite{danielewicz_effects_1995, andronic_systematics_2006}.  Our goal here, though, is of presenting deblurring opportunities and not of making of making everything optimal - actual experimental analysis will likely face other tensions.  In addition to the distribution found in the simulations, we show in Fig.~\ref{fig:dPhi} the central-limit result \eqref{eq:dPdPhi}, with widths from Eqs.~\eqref{eq:sigmas} and \eqref{eq:products}, and $N$ replaced by the average $\langle N \rangle$ for the simulation.  Remarkably, the curves cannot be distinguished by eye, which bodes well for using central-limit results in practice, in lieu of any complete self-consistent simulation.  Notably, testing of the proximity to the central-limit can be carried out directly in the experiment, by randomly dividing events into subevents of nearly equal multiplicity  \cite{danielewicz_transverse_1985} and comparing the relative distribution of estimated directions of the reaction plane to that from Eq.~\eqref{eq:dPdPhi} with $(N-1)$ replaced by $\langle N \rangle/4$.

\begin{figure}
\centering
\includegraphics[width=.55\linewidth]{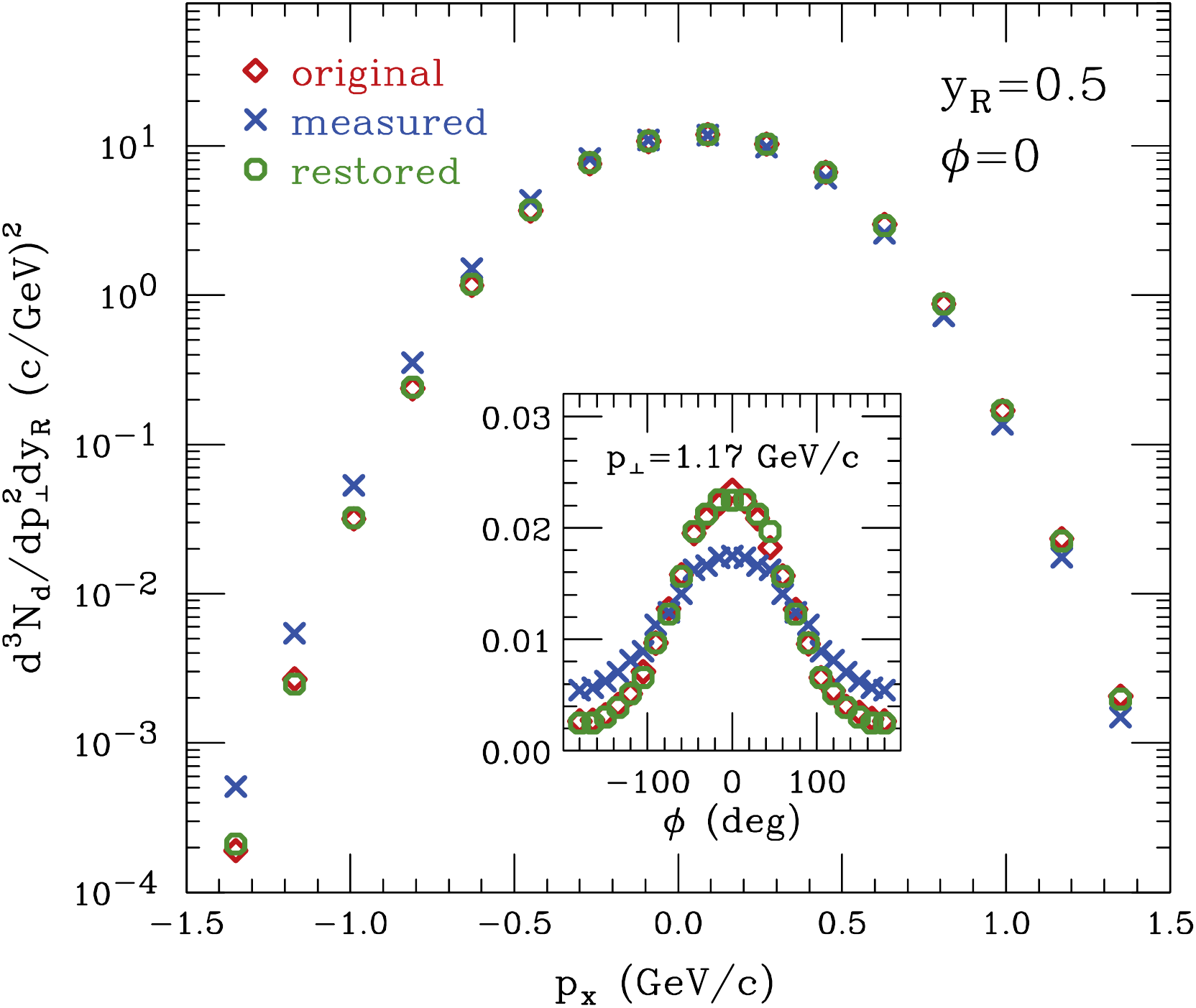}
\caption{Triple differential deuteron distribution at $y_R=0.5$, in the local equilibrium model.  The main figure shows the distribution in the reaction plane, as a function of the momentum component in the reaction plane.  The inset shows the distribution at $p_\perp=1.17 \, \text{GeV}/c$, as a function of azimuthal angle about the reaction plane.  In each case, three versions of the distribution are displayed: one determined directly using the known direction of the reaction plane in the modelled events (diamonds), one obtained by estimating the direction of the plane with the vector ${\bf Q}$  (crosses), out of momenta of remaining particles in the simulated events, Eq.~\eqref{eq:Q}, and one arrived by applying the deblurring to the measured triple-differential distribution, with the RL algorithm applied to Eq.~\eqref{eq:d3N} and the blurring function from the central limit theorem, Eq.~\eqref{eq:dPdPhi}.}
\label{fig:2dxd}
\end{figure}

We next illustrate the blurring that occurs when attempting to measure the triple differential distribution.  To the blurred distribution we apply deblurring.  Specifically, we examine the spectrum of deuterons at a moderately forward rapidity, $y_R=0.5$.  We choose deuterons rather than protons in the illustration, because the push of the latter in the reaction plane is relatively meager in the parameterization of the flows we chose.  Figure \ref{fig:2dxd} shows three versions of the transverse momentum spectra in the reaction plane ($p_y=0$): as recorded for the event sample using the known direction of the reaction plane (diamonds), as recorded (crosses) when using the direction estimated with a construct from particle momenta, Eqs.~\eqref{eq:Q} and \eqref{eq:qmu}, and as obtained from deblurring the blurred spectrum (circles).  The blurring function for the latter could be determined self-consistently, through additional simulations during the deblurring iterations, but given Fig.~\ref{fig:dPhi} we instead use the central-limit formula~\eqref{eq:dPdPhi}.  The discretization step in the RL algorithm is the same for the blurring function and the distribution and we use $H= 15\degree $.  The values from restoration settle after about 5 RL iterations with $\nu=1.99$ and $\lambda=0.005$.  The deblurring is carried out for each transverse momentum bin separately and the inset in Fig.~\ref{fig:2dxd} shows the case of a high central bin momentum, $p_\perp = 1.17 \, \text{GeV}/c$, with statistical fluctuations evident.

It is apparent in Fig.~\ref{fig:2dxd} that the restored spectrum well coincides with the original, as the respective representing symbols largely overlap.  It is apparent, for the high momenta there, that the restoration can reproduce spectra varying by an order of magnitude over the azimuthal angle.  Much of the variation over angle is already there in the simulated measured spectrum, see the inset.  The restoration just improves details in that variation and does so mostly on a local scale in the azimuthal angle.  Have we chosen different weights for $\bf Q$ in Eq.~\eqref{eq:omega}, the blurring function in Fig.~\ref{fig:dPhi} could have been more narrow, and there would be less needed improvement left for the deblurring.

For practitioners of the analysis of heavy-ion collisions, it should be apparent that the considerations in the 1D model earlier are just veiled versions of the reaction-plane analysis, stripped of the dimension that plays no direct role in the restoration and stripped of the periodicity in the other dimension.  With this, the steps that must be taken to treat detector inefficiencies consistently with the rest of the reaction-plane analysis should be apparent from the preceding section.  Particle misidentifications would give rise to matrix elements in the transfer matrix coupling not only different particle identities, but also rapidities.  Interestingly, it might even be possible to handle a situation where the particles could only be identified on a statistical level \cite{danielewicz_collective_1988}.  

\subsection{Why 3D Characteristics?}

It may be asked why engage in deblurring, if procedures exist for determining azimuthal Fourier coefficients, as a functions of transverse momentum \cite{poskanzer_methods_1998, borghini_flow_2001, bilandzic_flow_2011, hades_collaboration_directed_2020}.  The lowest nonvanishing coefficients usually have a straightfoward physical meaning and may be easier to determine for low event statistics than refined distributions.  However, a multitude of coefficients of different orders, changing with rapidity and transverse momentum, combining information from different azimuthal directions, can be a challenge at operational level.  To illustrate how the view of reactions could be expanded by examining distributions that include azimuthal angle relative to the reaction plane, and potentially 3D distributions, we reach to the pBUU transport model \cite{danielewicz_determination_2000}.  The latter and other transport models have been extensively used to simulate a variety of heavy-ion collisions in the beam energy region from few tens of MeV/nucl to few GeV/nucl, describing different observables at quantitative and semi-quantitative level \cite{wolter_2021}.  With this predictions of such models might be considered pertinent, but as azimuthal and more broadly 3D spectra have not been accessible experimentally, they were normally not considered in the theory either.

\begin{figure}
\centering
\includegraphics[width=.55\linewidth]{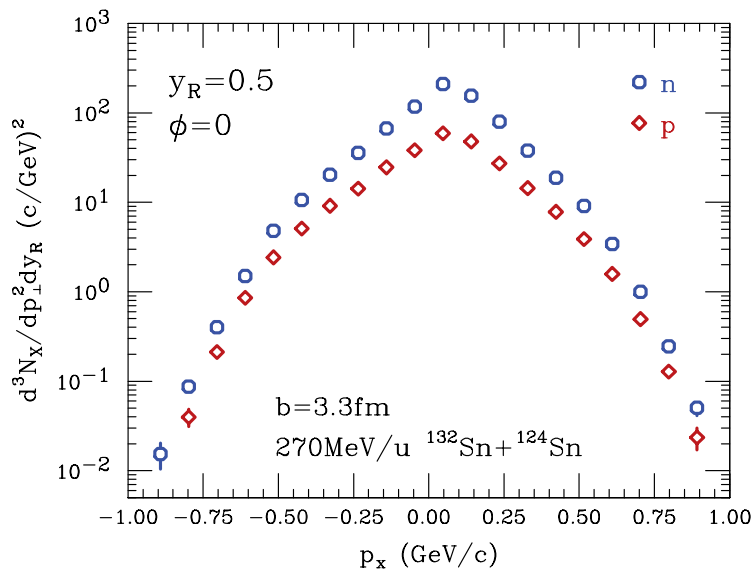}
\caption{Triple differential distributions in the reaction plane at $y_R=0.5$, for neutrons (circles) and protons (diamonds), from the pBUU simulations of $^{132}$Sn + $^{124}$Sn collisions at $270 \, \text{MeV/nucl}$ and $b=3.3 \, \text{fm}$.  The rapidity $y_R$ is here in the nucleon-nucleon center of mass and normalized to the beam.}
\label{fig:2dx2}
\end{figure}

Multiple transport simulations have been carried out in the context of $270 \, \text{MeV/nucl}$ Sn + Sn experiments performed recently at RIKEN \cite{jhang_symmetry_2021}.  In Fig.~\ref{fig:2dx2}, we show exemplary results from such simulations, proton and neutron spectra from $^{132}$Sn + $^{124}$Sn collisions at $b=3.3 \, \text{fm}$, within the reaction plane at $y_R=0.5$.  With projectile and target of different mass, the rapidity is taken here in the nucleon-nucleon center of mass and is normalized to the beam.  There are similarities and differences between the spectra in the conventional local equilibrium picture, such as in Fig.~\ref{fig:2dxd}, and from transport, such as in Fig.~\ref{fig:2dx2}, when the spectra are examined in the reaction plane.  Both types of spectra have maxima shifted in the positive direction of the reaction plane.  The one in the conventional picture is largely symmetric about the fairly flat maximum and close to parabolic in the logarithmic scale.  The maxima in the spectra from transport are sharp, the spectra are largely piecewise exponential in momentum, have knees and there is visible asymmetry between the sides extending into the positive and negative sides of the reaction plane.  The slope for the positive side, right after the maximum, is sharper than for the negative side.  This could be due to a larger fraction of spectator matter towards the positive side of the reaction plane in space, for particles moving forward, than towards the negative.  The knees in the spectra are at different transverse momenta on the two sides of the reaction plane.  A sharp maximum could be produced by Coulomb interactions, but these cannot explain the maximum for neutrons, that seems to be even sharper than for protons.

When the transport spectra are averaged over azimuthal angle, the maxima in the spectra move to $p_\perp=0$ and soften.  The knees soften too.  Overall, the two types of spectra, from transport and the conventional picture, become qualitatively much closer to each other with the azimuthal angle-averaging than without.  When azimuthal asymmetries in the spectra are explored in terms of Fourier coefficients, usually just one or two lowest ones in the particular energy regime, details such as in Fig.~\ref{fig:2dx2} are beyond resolution.

Differences in the spectra for the sides of the system with different participant-spectator composition on different sides of the reaction plane may be accessible only through the 3D examination of those spectra.  Shift of a maximum in the spectrum within the reaction plane may offer the only opportunity to examine its shape, as a maximum can be difficult to assess when it coincides with beam direction.  Comparison of the knees in relation to the maximum, between experiment and theory may help to clarify their origin and clarify the level of understanding of the collisions within theory, such as of the elementary collisions taking place on shell.

\begin{figure}
\centering
\includegraphics[width=.55\linewidth]{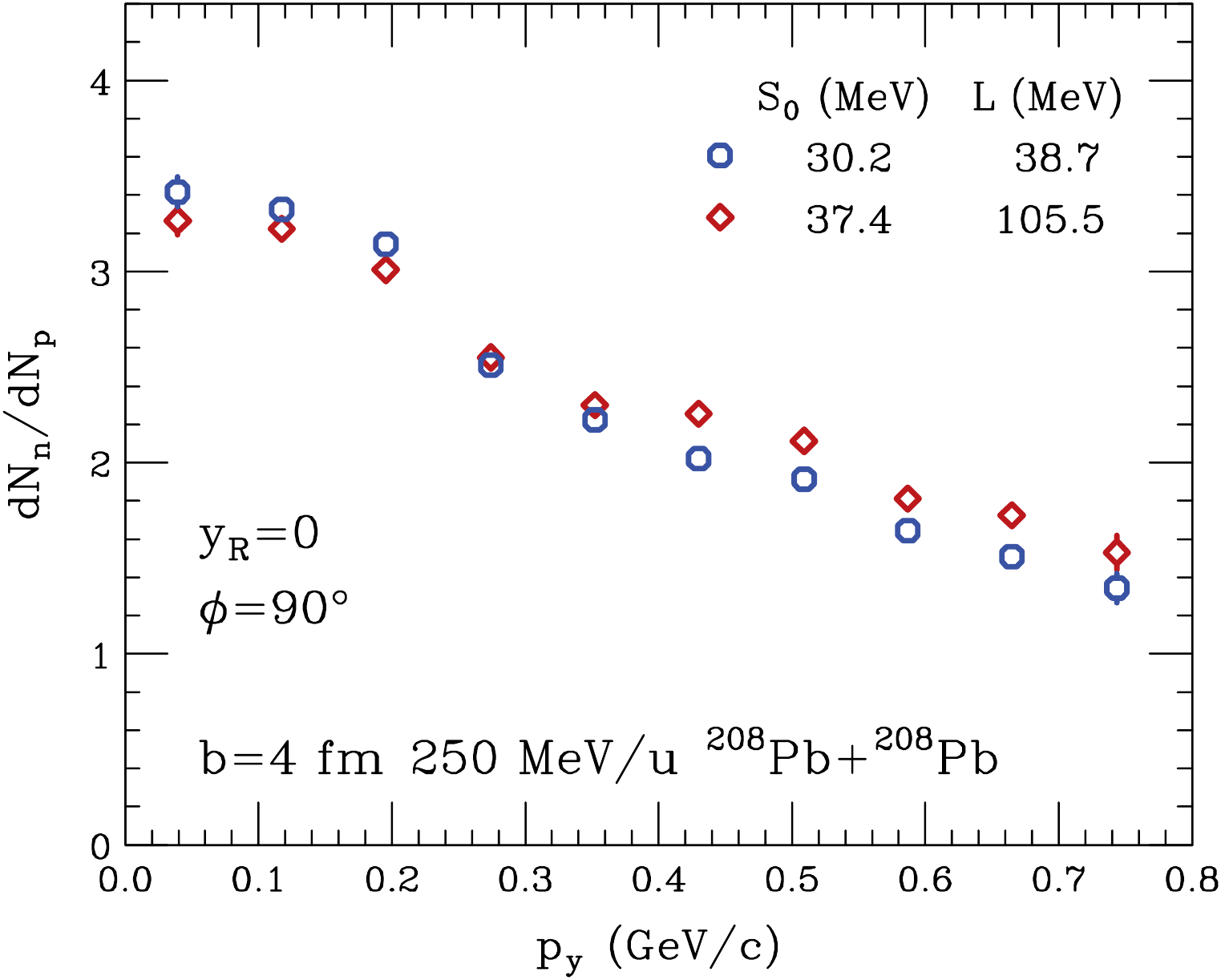}
\caption{Ratio of neutron yield to proton yield in $250 \, \text{MeV/nucl}$ $^{208}$Pb + $^{208}$Pb collisions at $b = 4 \, \text{fm}$, at midrapidity, in the direction perpendicular to the reaction plane, as a function of transverse momentum.  The two symbol sets represent the results for two different parametrizations of symmetry energy: diamonds - stiff with $L=105.5 \, \text{MeV}$ and circles - moderately soft with $L=38.7 \, \text{MeV}$.  The two parametrizations are chosen so that they yield the symmetry energy at $\rho \sim 2\rho_0/3$ consistent with the conclusions from binding energies of heavy nuclei and other determinations pertaining to subnormal densities.}
\label{fig:rnp}
\end{figure}

Turning to the second example, heavy-ion collisions compress nuclear matter to densities above normal at conditions approaching thermal equilibrium.  With this they represent opportunity to learn about the nuclear equation of state~(EOS).  The current interest is in the component of EOS, so-called symmetry energy that describes energy change with the change in relative neutron-proton asymmetry, at different net nucleonic densities.  Energies of more massive nuclei, and other data, constrain the symmetry energy at subnormal densities and especially at the density representing an average for the nuclei, $\rho \simeq 2\rho_0/3$, at a value of about $25.4 \, \text{MeV}$ \cite{lynch_decoding_2021}.  From there on, the region in density opens up where the symmetry energy is poorly constrained and this includes the pace of symmetry-energy variation around $\rho_0$.  Yet when different parametrizations of the symmetry energy are explored in transport simulations, that pass near the asserted value, it is very hard to find sensitivity to the symmetry energy at $\rho \gtrsim \rho_0$, in nearly any predicted observable.  One reason are low asymmetries for the nuclear systems as a whole, additionally depleted in the center of the matter due to migration of the asymmetry to the surface.  There is some advantage in going to heavier systems, as surface to volume ratio drops and Coulomb interactions help to maintain significant interior asymmetry, both in the initial state and in reaction dynamics \cite{stone_2021}.  Access to the 3D information should help too.  Indeed, if one examines nucleonic spectra in direction perpendicular to the reaction plane, in $250 \, \text{MeV/nucl}$ $^{208}$Pb + $^{208}$Pb collisions at $b=4 \, \text{fm}$, at $y_R=0$, see Fig.~\ref{fig:rnp}, one can observe definite changes in the ratio of neutron to proton yields, when the symmetry energy evolves from moderately soft to stiff, with its slope parameter at $\rho_0$ changing from $L=38.7 \, \text{MeV}$ to $L=105.5 \, \text{MeV}$.  Both parametrizations of the symmetry pass by the consensus value at $\rho \sim 2\rho_0/3$.

The fact that a stiff symmetry energy raises the neutron-proton ratio at high transverse momentum agrees with intuitive expectations.  However, the fact that it lowers the ratio at low momenta is somewhat surprising.   Likely the unexpected features of both Figs.~\ref{fig:2dx2} or \ref{fig:rnp} are tied to evolution of particle emission with time.  The ratio being higher than the overall for the system of 1.54 in Fig.~\ref{fig:rnp}, for most transverse momenta, irrespective of the EOS, may be due to Coulomb interactions that push protons out not only in the transverse directions, but also along the beam axis. 

In practical comparisons of transport theory to data it is common to make sure that many more rudimentary aspects of the data are understood on equal footing with those more sophisticated.  I.e., other aspects of measured and calculated distributions would need to be simultaneously checked and more calculations would need to be carried out, than here, to draw credible physics conclusions. 

\section{Conclusions}

We have proposed to apply deblurring, adopted from optics, to nuclear observables subject to degradation, whether on account of the inference method or detector performance, such as 3D characteristics relative to the reaction plane in heavy-ion collisions.  In some instances, such as that of the reaction plane determination or the schematic 1D model, and sufficiently many particles detected, the blurring function may be determined, relatively faithfully, following the central limit theorem. In a more general situation, the blurring function may need to be determined through simulations, possibly carried out self-consistently with the deblurring, as for the 1D model with inefficiency that cannot be linearized, or through independent measurements, as in assessing detector performance.  In heavy-ion collisions, the deblurring, such as with the Richardson-Lucy method, can work locally in the azimuthal angle only improving the 3D characteristics of collision events that are already resolved to some extent with the estimated reaction plane.  This is a potential improvement over the complete Fourier decomposition of 3D characteristics in the azimuth.  Importantly, estimation of the reaction plane can be made with particles measured in a different detector than the one used to detect the particle or particles for which the 3D characteristics are desired.

In analysis of actual heavy-ion data, other correlations, than those associated with the reaction plane, will be present in the final states, in particular tied to the total momentum conservation, interactions at low relative velocity and sequential decays.  Strategies to deal with these have been developed in other contexts \cite{danielewicz_collective_1988, borghini_effects_2002, hades_collaboration_directed_2020} and first the base needs to be established ignoring these correlations, which has been attempted here.

Finally, we hope that the deblurring, revealing 3D distributions, could help extract novel physics information from the heavy-ion collisions, that we tried to illustrate.  Beyond single-particle distributions tied to the reaction plane, we hope that the strategy could be used for low-velocity correlations \cite{lisa_azimuthal_2000}.

\begin{acknowledgements}
The authors benefited from discussions with Scott Pratt and with members of the S$\pi$RIT Collaboration.
This work was supported by the U.S.\ Department of Energy Office of Science under Grant {DE}-{SC}0019209.

\end{acknowledgements}

\bibliography{deblur}

\end{document}